\documentstyle[12pt]{article}
\newlength{\mathspace}
\def\h #1{\hat{#1}}
\def\b #1{\bar{#1}}
\def\np#1{ Nucl. Phys. B#1}
\def\pr#1    { Phys. Rev. D#1 }
\def\pl#1{ Phys. Lett. B#1}

\def\ijmp#1  { Int. Jour. Mod. Phys. A#1 }
\def\mpl#1   { Mod. Phys. Lett. A#1 }
\def\begineq{\begin{equation}}
\def\endeq{\end{equation}}
\def\eqabegin{\begin{eqnarray}}
\def\eqaend{\end{eqnarray}}
\def\nn{\nonumber}

\begin{document}
\baselineskip=0.7cm
\setlength{\mathspace}{2.5mm}



\begin{titlepage}

    \begin{normalsize}
     \begin{flushright}
                 SINP-TNP/97-02\\
                 hep-th/9705016\\
     \end{flushright}
    \end{normalsize}
    \begin{LARGE}
       \vspace{1cm}
       \begin{center}
         {On S-Duality of Toroidally Compactified}\\ 
         {Type IIB String Effective Action}\\ 
       \end{center}
    \end{LARGE}

  \vspace{5mm}

\begin{center}
           
             \vspace{.5cm}

            Shibaji R{\sc oy}
           \footnote{E-mail address: roy@tnp.saha.ernet.in} 

                 \vspace{2mm}

{\it Saha Institute of Nuclear Physics}\\
        {\it 1/AF Bidhannagar, Calcutta 700 064, India}\\

      \vspace{2.5cm}

    \begin{large} ABSTRACT \end{large}
        \par
\end{center}
 \begin{normalsize}
\ \ \ \
It has been shown recently that the toroidally compactified type IIB string 
effective action possesses an SL(2, R) invariance as a consequence of the
corresponding symmetry in ten dimensions when the self-dual five-form field
strength is set to zero. By working in the string frame
we clarify how a Z$_2$ subgroup of this SL(2, R)
group responsible for producing the strong-weak coupling duality in
the ten dimensional theory produces the same symmetry for the reduced theory.
In the
absence of the full covariant action of type IIB supergravity theory, we
show that the T-dual version of type IIA string effective action (including
the R-R terms) in D=9 also possesses the SL(2, R) invariance indicating that
this symmetry is present for the full type IIB string effective action 
compactified on torus.

\end{normalsize}

\end{titlepage}
\vfil\eject

By now there is a mounting evidence in support of the conjecture that type
IIB superstring theory in ten dimensions has an SL(2, R) invariance [1--3]. 
The
discrete subgroup of this SL(2, R) group survives as an exact symmetry of 
the quantum theory and has been referred to as S-duality in the literature
in analogy with the corresponding symmetry in N=4, D=4 heterotic string
theory [4,5]. Under this SL(2, R) transformation 
the ten dimensional dilaton of
type IIB string theory transforms non-trivially and so it mixes up the
different orders of the string perturbation theory. In particular, a 
Z$_2$ subgroup of this SL(2, R) group relates the weak and the strong
coupling regime of the type IIB string theory in D=10, when the scalar field
in the Ramond-Ramond (R-R) sector of the spectrum is set to zero. Thus this
symmetry is non-perturbative and that is why it is difficult to prove this
conjecture. Many interesting consequences of this symmetry have been explored
in refs.[3,6,7]. In particular, the various classical p-brane 
solutions of type
IIB string theory have been shown to form SL(2, Z) multiplets and as a
consequence the existence of bound states of $n$ fundamental p-branes with
$m$ Dirichlet p-branes with $(n,\,m)$ relatively prime integers, has been
predicted in ref.[8]. Although it is difficult to prove this conjecture,
there are various indirect ways to understand the origin of this symmetry.
Purely, from string theory point of view the SL(2, R) symmetry can be 
understood from the ten-dimensional string-string duality conjecture of
Type I and heterotic string theory with gauge group SO(32) and T-duality
[2,9].
This symmetry can also be understood from the hypothetical higher dimensional
theories called `M-theory' [6,10] and `F-theory' [11] compactified on torus.

In this paper, we will study the toroidal compactification [12,13] of 
type IIB string
theory and explore some of the consequences of SL(2, R) invariance of ten
dimensional theory. Since the ten dimensional theory already has SL(2, R)
invariance, it is expected that this symmetry will persist also in lower
dimensions. We will first show that it is indeed the case. It is well-known
[14,15] that the full type IIB 
supergravity theory is non-Lagrangian because of the    
presence of a self-dual four-form gauge field in the spectrum. But, once
we set the corresponding field strength to zero, a consistent covariant
action can be written from which the field equations of type IIB supergravity
theory could be obtained [2]. We will take this type 
IIB string effective action
and reduce it on (10 $-$ D) dimensional torus in the string frame. This
D-dimensional effective action when written down in the Einstein frame
will be shown to possess an SL(2, R) invariance as expected. The toroidal
compactification of the same type IIB  string effective action in the
Einstein frame has been studied recently by Maharana [16] and the SL(2, R)
invariance of the reduced action was shown to follow directly in this way.
We will point out that in showing this invariance of the toroidally 
compactified theory, it is not the D-dimensional dilaton but rather a
linear combination of the dilaton and other moduli of the theory which 
appears in
the matrix ${\cal M}_{\rm D}$ (see eq.(19) and the 
discussion after eq.(21)).
It is for this reason, a Z$_2$ subgroup of the SL(2, R) group which is
responsible for producing the strong-weak coupling duality symmetry in
the ten dimensional theory does not necessarily imply the 
same symmetry in the reduced 
theory. We will clarify how the strong-weak coupling duality symmetry can be
understood in the reduced theory by working in the string frame. It should
be emphasized that this effect becomes
transparent if we reduce the action in the string frame and then go over
to the Einstein frame rather than reducing the theory directly in the 
Einstein frame
as in ref.[16]. Then we will show that not only this truncated version of the
type IIB string effective action but also the full action has the SL(2, R)
invariance when compactified on torus. In the absence of the full action, we
will take the following strategy to understand this symmetry in the full
theory. Since the complete type IIA string effective action including the
R-R terms is known [17,18], we reduce this action 
on a circle. It should be pointed 
out here that it is difficult to study the toroidal compactification of
this action in general because of the presence of the topological terms 
whose forms are very specific to the dimensionality of the reduced theory.
We will therefore look at the simplest case i.e. the reduction on a circle.
Since in D=9, type IIA and type IIB theories are T-dual to each other
[19,20], we will
take the T-dual version of this D=9 type IIA action and then show that this 
action is indeed SL(2, R) invariant. Thus we conclude that the full type IIB
string effective action when toroidally compactified also possesses an SL(2, R) 
invariance.

Let us recall that the massless spectrum of the type IIB string theory in 
the bosonic sector contains a graviton $\h g_{B,\, \h \mu \h \nu}$, a
dilaton $\h \phi^{(1)}$ and an antisymmetric tensor field $\h b_{\h \mu
\h \nu}^{(1)}$ as Neveu-Schwarz--Neveu-Schwarz (NS-NS) sector states 
whereas in the R-R sector it contains another scalar $\h {\phi}^{(2)}$,
another antisymmetric tensor field $\h b_{\h \mu \h \nu}^{(2)}$ and a
four-form gauge field $\h A_{\h \mu \h \nu \h \rho \h \sigma}^+$ whose
field strength is self-dual [21]\footnote[1]{We denote the ten dimensional
space-time coordinates and the fields with a `hat'. The objects in lower
dimensions will be denoted without `hat'.}. When the field strength
associated with the four-form gauge field is set to zero, the type IIB
supergravity equation of motion can be obtained from the following covariant
action [2]:
\eqabegin
S_{{\rm IIB}}^{(10)} &=&
\int\,d^{10} \h x \sqrt {- \h {g}_B} \left[ e^{-2 \h {\phi}^{(1)}}
\left(\h {R}_B + 4 \partial_{\h \mu} \h {\phi}^{(1)} \partial^{\h \mu} 
\h {\phi}^{(1)}
- \frac{1}{12} \h {h}_{\h \mu \h \nu \h \lambda}^{(1)} \h {h}^{(1)\, \h \mu
\h \nu \h \lambda}\right)\right.\nn\\
& &\qquad \left. - \frac{1}{2} \partial_{\h \mu} \h {\phi}^{(2)} \partial^
{\h \mu}
\h {\phi}^{(2)} -\frac{1}{12}\left(\h {h}_{\h \mu \h \nu \h \lambda}^{(2)} 
+ \h {\phi}^{(2)} \h {h}_{\h \mu \h \nu \h \lambda}^{(1)}\right)\left(
\h {h}^{(2)\, \h \mu \h \nu \h \lambda} + \h {\phi}^{(2)} \h {h}^{(1)\,
\h \mu \h \nu \h \lambda}\right)\right]
\eqaend
where
\begineq
\h {h}_{\h \mu \h \nu \h \lambda}^{(i)}\,\,=\,\,\left(\partial_{\h \mu}
\h {b}_{\h \nu \h \lambda}^{(i)} + {\rm cyc.\,\,\, in\,\,\, \h \mu \h \nu
\h \lambda}\right), \qquad i=1,2
\endeq
The action in (1) is known to possess a global SL(2, R) invariance which can 
be better understood in the Einstein frame where the Einstein metric and the
string metric are related as $\b {\h g}_{B,\,\h \mu \h \nu} = e^{- \frac{1}{2}
\h {\phi}^{(1)}} \h {g}_{B,\,\h \mu \h \nu}$. In the Einstein frame the action 
can be written as:
\eqabegin
\b {S}_{{\rm IIB}}^{(10)} &=&
\int\,d^{10} \h x \sqrt {- \b {\h g}_B} \left[ \b {\h R}_B - \frac {1}{2}
\partial _{\h \mu} \h {\phi}^{(1)} \partial ^{\h \mu} \h {\phi}^{(1)} -\frac
{1}{2} e^{2 \h {\phi}^{(1)}} \partial_{\h \mu} \h {\phi}^{(2)} \partial ^{\h
\mu}\h {\phi}^{(2)}\right.\nn\\
& &\left. -\frac{1}{12}\left(e^{-\h {\phi}^{(1)}} \h {h}_{\h \mu \h \nu \h \lambda}
^{(1)} \h {h}^{(1)\,\h \mu \h \nu \h \lambda} + e^{\h {\phi}^{(1)}} \left(
\h {h}_{\h \mu \h \nu \h \lambda}^{(2)} + \h {\phi}^{(2)} \h {h}_{\h \mu \h
\nu \h \lambda}^{(1)}\right)\left(\h {h}^{(2)\, \h \mu \h \nu \h \lambda} +
\h {\phi}^{(2)} \h {h}^{(1)\,\h \mu \h \nu \h \lambda}\right)\right)\right]
\eqaend
This action can now be expressed in a manifestly SL(2, R) invariant form
as given below:
\begineq
\b {S}_{{\rm IIB}}^{(10)}\,\,=\,\, \int\,d^{10} \h x \sqrt {- \b {\h g}_B}
\left[\b {\h R}_B + \frac{1}{4} {\rm tr}\, \partial_{\h \mu} \h {\cal M}
\partial^{\h \mu}\h {\cal M}^{-1} - \frac{1}{12}\h {\bf h}_{\h \mu \h \nu
\h \lambda}^T \h {\cal M} \h {\bf h}^{\h \mu \h \nu \h \lambda}\right]
\endeq 
where, 
\begineq
\h {\cal M}\,\,\equiv\,\, \left(\begin{array}{cc}
\left(\h {\phi}^{(2)}\right)^2 e^{\h {\phi}^{(1)}} + e^{- \h {\phi}^{(1)}}
& \h {\phi}^{(2)} e^{\h {\phi}^{(1)}} \\
\h {\phi}^{(2)} e^{\h {\phi}^{(1)}}  &  e^{\h {\phi}^{(1)}}\end{array}\right)
\endeq
represents an SL(2, R) matrix and $\h {\bf h}_{\h \mu \h \nu \h \lambda} \,\,
\equiv \,\, \left(\begin{array}{c} \h {h}_{\h \mu \h \nu \h \lambda}^{(1)}  \\
\h {h}_{\h \mu \h \nu \h \lambda}^{(2)}\end{array}\right)$. Also, the 
superscript `$T$' represents the transpose of a matrix.
The action (4) can be easily seen to be invariant under a global SL(2, R)
transformation $\h {\cal M} \rightarrow \Lambda \h {\cal M} \Lambda^T$ and
$\left(\begin{array}{c} \h {b}_{\h \mu \h \nu}^{(1)} \\  \h {b}_{\h \mu \h \nu
}^{(2)}\end{array}\right)\,\,\equiv\,\, \h {\bf b}_{\h \mu \h \nu} 
\rightarrow (\Lambda^{-1})^T \h {\bf b}_{\h \mu \h \nu}$, where $\Lambda\,\,
=\,\,\left(\begin{array}{cc} a & b \\ c & d\end{array}\right)$, with $ ad - bc
= 1$, represents a global SL(2, R) transformation matrix. Note that the 
canonical metric $\b {\h g}_{B\,\h \mu \h \nu}$ remains invariant under the
SL(2, R) transformation. With these transformations the complex scalar field
$\h \rho = \left(\h {\phi}^{(2)} + i e^{-\h {\phi}^{(1)}}\right)$ undergoes
 a fractional linear transformation whereas the two two-form potentials
$\h {b}_{\h \mu \h \nu}^{(1)}$ and $\h {b}_{\h \mu \h \nu}^{(2)}$ transform
linearly [3]. 
In particular, choosing $\Lambda = \left(\begin{array}{cc} 0 & 1 \\
-1 & 0 \end{array}\right)$ and $\h {\phi}^{(2)} = 0$, the string coupling
constant $e^{\h {\phi}^{(1)}}$ transforms to its inverse showing a 
strong-weak coupling duality symmetry in the theory. 

We now perform the dimensional reduction of the action (1) on (10 $-$ D)
dimensional torus. In order to achieve this we first split the 10-dimensional
coordinates $\h {x}^{\h \mu}$ to D space-time coordinates
$x^\mu$ and (10 $-$ D)
internal coordinates $x^m$ and demand that all the D-dimensional fields be
independent of the compact coordinates
[12,13]. By using the Lorentz
invariance the ten dimensional vielbein is usually taken in the following
triangular form
[22]\footnote[2]{Here the Greek letters $(\lambda,\,\mu,\,\ldots)$
in the later part of the alphabet denote the curved space-time indices whereas
$(\alpha,\,\beta,\,\ldots)$ in the beginning of the alphabet correspond to flat
tangent space indices. Similarly, the Latin letters $(m,\,n,\,\ldots)$
represent the internal indices and $(a,\,b,\,\ldots)$ denote the corresponding
tangent space indices.}
\begineq
\h {e}^{\h \alpha}_{\h \mu}\,\,=\,\,\left(\begin{array}{cc} e_{\mu}^{\alpha}
& a_{\mu}^{(3)\,n} e_n^a  \\
0 & e_m^a\end{array}\right)
\endeq
where $e_\mu^\alpha$ and $e_m^a$ are respectively the space-time and internal
vielbeins. $a_\mu^{(3)\,n}$ are (10 $-$ D) vector gauge fields resulting
from the dimensional reduction of the metric. (We have reserved $a_\mu^{(1)
\,n}$ and $a_\mu^{(2)\,n}$ to denote the gauge fields resulting from the
dimensional reduction of the two antisymmetric tensor fields $\h {b}
_{\h \mu \h \nu}^{(1)}$ and $\h {b}_{\h \mu \h \nu}^{(2)}$.) The 
D-dimensional metric and the internal metric are given respectively by
$g_{B,\,\mu\nu} = e_\mu^\alpha e_\nu ^\beta \eta_{\alpha \beta}$ and 
$g_{mn} = e_m^a e_n^b \delta_{ab}$, where our convention for the signature
of the Lorentz metric is $(-, +, +, \cdots)$. Note that with this convention
(6), $\sqrt{ - \h {g}_B} = \sqrt{- g_D}\,\, \Delta$, where $\h {g}_B = 
({\rm det}\,
\, \h {g}_{B,\,\h \mu \h \nu})$, $g_D = ({\rm det}\,\, g_{D,\,\mu\nu})$ and
$\Delta^2 = ({\rm det}\,\, g_{mn})$. With these definitions the scalar
part of the action (1) reduces to [18]:
\eqabegin
& &\int\,d^{10} \h x \sqrt {- \h {g}_B} \left[ e^{-2 \h {\phi}^{(1)}}
\left(\h {R}_B + 4 \partial_{\h \mu} \h {\phi}^{(1)} \partial^{\h \mu} 
\h {\phi}^{(1)}
\right)
 - \frac{1}{2} \partial_{\h \mu} \h {\phi}^{(2)} \partial^
{\h \mu}
\h {\phi}^{(2)}\right]\nn\\
&\longrightarrow& \int\, d^D x \sqrt{-{g_D}}\left[e^{-2 \phi^{(1)}_D} 
\left(
R_D + 4 \partial_\mu \phi^{(1)}_D \partial^\mu \phi^{(1)}_D - \frac{1}{4}
g_{mn} f_{\mu\nu}^{(3)\,m} f^{(3)\,\mu\nu,\,n}\right.\right.\nn\\
& &\qquad\qquad\qquad\left.\left. + \frac{1}{4} 
\partial_\mu g_{mn} \partial^\mu g^{mn}
\right) - \frac{1}{2} \Delta \partial_\mu \phi^{(2)} \partial^\mu \phi^{(2)}
\right]
\eqaend
where $R_D$ is the scalar curvature for the D-dimensional metric 
$g_{D,\,\mu\nu}$, $\h {\phi}^{(1)} = \phi_D^{(1)} + \frac{1}{2} \log \Delta$,
$\phi_D^{(1)}$ being the D-dimensional dilaton and $\h {\phi}^{(2)} = 
\phi^{(2)}$. Also, $f_{\mu\nu}^{(3)\,m} = \partial_\mu a_\nu^{(3)\,m} -
\partial_\nu a_\mu^{(3)\,m}$. The reduction of the terms involving square of
the field strengths associated with the antisymmetric tensor fields $\h {b}
_{\h \mu \h \nu}^{(i)}$ in the action (1) is given below:
\eqabegin
& &-\frac{1}{12}\int\,d^{10} 
\h x \sqrt {- \h {g}_B} \left[ e^{-2 \h {\phi}^{(1)}}
 \h {h}_{\h \mu \h \nu \h \lambda}^{(1)} \h {h}^{(1)\, \h \mu
\h \nu \h \lambda}
 +\left(\h {h}_{\h \mu \h \nu \h \lambda}^{(2)} 
+ \h {\phi}^{(2)} \h {h}_{\h \mu \h \nu \h \lambda}^{(1)}\right)\left(
\h {h}^{(2)\, \h \mu \h \nu \h \lambda} + \h {\phi}^{(2)} \h {h}^{(1)\,
\h \mu \h \nu \h \lambda}\right)\right]\nn\\
&\longrightarrow& \int\, d^D x \sqrt{-{g_D}}
\left[e^{-2\phi_D}\left(-\frac{1}{4}
g^{mp} g^{nq}\partial_\mu b_{mn}^{(1)} \partial^\mu b_{pq}^{(1)} -\frac{1}{4}
g^{mp} h_{\mu\nu\,m}^{(1)} h^{(1)\,\mu\nu}_{\,\,\,\,p} - \frac{1}{12}
h_{\mu\nu\lambda}^{(1)} h^{(1)\,\mu\nu\lambda}\right)\right.\nn\\
& &\qquad\qquad\qquad -\frac{1}{4} \Delta g^{mp} g^{nq}\left(\partial_\mu 
b_{mn}^{(2)} + \phi^{(2)} \partial_\mu b_{mn}^{(1)}\right)\left(\partial^\mu
b_{pq}^{(2)} + \phi^{(2)} \partial^\mu b_{pq}^{(1)}\right)\nn\\
& &\qquad\qquad\qquad\qquad -\frac{1}{4}\Delta g^{mp}
\left(h_{\mu\nu\,m}^{(2)} + \phi^{(2)}
h^{(1)}_{\mu\nu\,m}\right)\left(h^{(2)\,\mu\nu}_{\,\,\,\,p} + \phi^{(2)}
h^{(1)\,\mu\nu}_{\,\,\,\,p}\right)\nn\\
& &\qquad\qquad\qquad\qquad\left. -\frac{1}{12}
\Delta\left(h^{(2)}_{\mu\nu\lambda} +
\phi^{(2)} h^{(1)}_{\mu\nu\lambda}\right)\left(h^{(2)\,\mu\nu\lambda} +
\phi^{(2)} h^{(1)\,\mu\nu\lambda}\right)\right]
\eqaend
Our definitions and the reduced form of various gauge fields are:
\eqabegin
\h g_{B,\, \h \mu \h \nu} &\longrightarrow& \left\{\begin{array}{l}
\h g_{B,\,mn}\,\,=\,\,g_{mn}\\
g_{B,\, \mu m}\,\,=\,\,\h g_{B,\,\mu m}\,\,=\,\, a_\mu^{(3)\,n} g_{mn}\\
\h g_{B,\,\mu\nu}\,\,=\,\, g_{B,\,\mu\nu} + g_{mn} a_\mu^{(3)\,m} a_{\nu}
^{(3)\,n}\end{array}\right.\\
\h {\phi}^{(1)}\,\,&=&\,\, \phi_D^{(1)} + \frac{1}{2} \log \Delta\\
\h \phi^{(2)}\,\,&=&\,\, \phi^{(2)}\\
\h {b}_{\h \mu \h \nu}^{(i)} &\longrightarrow& \left\{\begin{array}
{l} b_{\mu m}^{(i)}\,\,=\,\, a_{\mu m}^{(i)}\,\,=\,\, \h {b}_{\mu m}^{(i)}
+ b_{mn}^{(i)} a_{\mu}^{(3)\,n}\\
b_{\mu\nu}^{(i)}\,\,=\,\, \h {b}_{\mu\nu}^{(i)} + a_\mu^{(3)\,m} 
a_{\nu m}^{(i)} - a_\nu^{(3) m} a_{\mu m}^{(i)} - a_\mu^{(3) m} b_{mn}^{(i)}
a_\nu^{(3) n}\end{array}\right.
\eqaend
The corresponding field strengths are given as,
\eqabegin
h_{\mu mn}^{(i)} &=& \h {h}_{\mu mn}^{(i)}\,\,=\,\,\partial_\mu b_{mn}^{(i)}\\
h_{\mu\nu m}^{(i)} &=& f_{\mu\nu m}^{(i)} - b_{mn}^{(i)} f_{\mu\nu}^{(3)\,n}
\eqaend
where 
\begineq
f_{\mu\nu m}^{(i)}\,\,=\,\, \partial_\mu a_{\nu m}^{(i)} - \partial_\nu
a_{\mu m}^{(i)}
\endeq
and finally,
\begineq
h_{\mu\nu\lambda}^{(i)}\,\,=\,\,\partial_\mu b_{\nu\lambda}^{(i)} - f_{\mu\nu
}^{(3)\,m} a_{\lambda\,m}^{(i)} + {\rm cyc.\,\,\, in\,\,\, \mu\nu\lambda
}\endeq
It is well-known that the NS-NS sector of the action (1) possesses a global
non-compact symmetry O(10 $-$ D, 10 $-$ D) when compactified on (10 $-$ D)
dimensional torus [23]. In order to show this invariance it was pointed out
in ref.[18], that it is necessary to express $h_{\mu\nu\lambda}^{(1)}$ in
(16) in terms of $\b b_{\mu\nu}^{(1)}\,\,\equiv\,\, b_{\mu\nu}^{(1)} -
\frac{1}{2} \left(a_\mu^{(3)\,m} a_{\nu\,m}^{(1)} - a_{\nu}^{(3)\,m} 
a_{\mu\,m}^{(1)}\right)$ as this remains invariant under O(10 $-$ D, 
10 $-$ D) transformation. In the present context, the introduction of
`barred' fields is not necessary.
 So, the
complete form of the dimensionally reduced action (1) to D-dimensions is
given as:
\eqabegin
& & \int\, d^D x \sqrt{-{g_D}}\left[e^{-2 \phi^{(1)}_D} 
\left(
R_D + 4 \partial_\mu \phi^{(1)}_D \partial^\mu \phi^{(1)}_D - \frac{1}{4}
g_{mn} f_{\mu\nu}^{(3)\,m} f^{(3)\,\mu\nu,\,n} + \frac{1}{4}
\partial_\mu g_{mn}\partial^\mu g_{mn}\right.\right.\nn\\
& &\qquad\qquad\left. 
-\frac{1}{4}
g^{mp} g^{nq}\partial_\mu b_{mn}^{(1)} \partial^\mu b_{pq}^{(1)}
  -\frac{1}{4}
g^{mp} h_{\mu\nu\,m}^{(1)} h^{(1)\,\mu\nu}_{\,\,\,\,p} - \frac{1}{12}
h_{\mu\nu\lambda}^{(1)} h^{(1)\,\mu\nu\lambda}\right)\nn\\
& &\qquad\qquad -\frac{1}{2} \Delta \partial_\mu \phi^{(2)} \partial^\mu
\phi^{(2)}
 -\frac{1}{4} \Delta g^{mp} g^{nq}\left(\partial_\mu 
b_{mn}^{(2)} + \phi^{(2)} \partial_\mu b_{mn}^{(1)}\right)\left(\partial^\mu
b_{pq}^{(2)} + \phi^{(2)} \partial^\mu b_{pq}^{(1)}\right)\nn\\
& &\qquad\qquad\qquad -\frac{1}{4}\Delta g^{mp}\left(h_{\mu\nu\,m}
^{(2)} + \phi^{(2)}
h^{(1)}_{\mu\nu\,m}\right)\left(h^{(2)\,\mu\nu}_{\,\,\,\,p} + \phi^{(2)}
h^{(1)\,\mu\nu}_{\,\,\,\,p}\right)\nn\\
& &\qquad\qquad\qquad\left. -\frac{1}{12}\Delta\left(h^{(2)}_{\mu\nu\lambda} +
\phi^{(2)} h^{(1)}_{\mu\nu\lambda}\right)\left(h^{(2)\,\mu\nu\lambda} +
\phi^{(2)} h^{(1)\,\mu\nu\lambda}\right)\right]
\eqaend
We have obtained the reduced action in the string frame metric and the
D-dimensional string coupling constant is given by $\lambda_D \sim e^
{\phi_D^{(1)}}$, where $\phi_D^{(1)}$ is the D-dimensional dilaton whose
relation with the ten dimensional dilaton is given in eq.(10). Just like the 
ten dimensional case, the SL(2, R) invariance of the action (17) becomes
manifest in the Einstein frame. In order to rewrite the action in the Einstein
frame we rescale the string metric as usual by $\b {g}_{D,\, \mu\nu}\,=\,
e^{-\frac{4}{D-2} \phi_D^{(1)}} g_{D,\,\mu\nu}$. Then the action (17) takes
the following form:
\eqabegin
& & \int\, d^D x \sqrt{-\b{g}_D}\left[ 
\b{R}_D -\frac{1}{2} \partial_\mu \b{\phi}^{(1)}_D \partial^\mu 
\b{\phi}^{(1)}_D -\frac{1}{2} e^{2 \b{\phi}^{(1)}_D} \partial_\mu \phi^{(2)}
\partial^\mu \phi^{(2)} + \frac{1}{8} \partial_\mu \log {\b {\Delta}}
\partial^\mu \log{\b{\Delta}}\right.\nn\\ 
& &\qquad +\frac{1}{4} \partial_\mu \b {g}_{mn} \partial^\mu \b{g}^{mn}
- \frac{1}{4}
\b{g}_{mn} f_{\mu\nu}^{(3)\,m} f^{(3)\,\mu\nu,\,n}
-\frac{1}{4}(\b {\Delta})^{1/2}
\b{g}^{mp} \b{g}^{nq} e^{-\b{\phi}^{(1)}_D}\partial_\mu b_{mn}^{(1)} 
\partial^\mu b_{pq}^{(1)}\nn\\
 & &\qquad
 -\frac{1}{4}(\b{\Delta})^{1/2}\b {g}^{mp} \b{g}^{nq} e^{\b {\phi}^{(1)}_D}
\left(\partial_\mu 
b_{mn}^{(2)} + \phi^{(2)} \partial_\mu b_{mn}^{(1)}\right)\left(\partial^\mu
b_{pq}^{(2)} + \phi^{(2)} \partial^\mu b_{pq}^{(1)}\right)\\
& &\qquad -\frac{1}{4}(\b{\Delta})^{1/2}\b {g}^{mp}\left
\{e^{-\b{\phi}^{(1)}_D}
h_{\mu\nu m}^{(1)} h^{(1)\,\mu\nu}_{\,\,\,\,p} + e^{\b{\phi}^{(1)}_D}
\left(h_{\mu\nu\,m}^{(2)} + \phi^{(2)}
h^{(1)}_{\mu\nu\,m}\right)\left(h^{(2)\,\mu\nu}_{\,\,\,\,p} + \phi^{(2)}
h^{(1)\,\mu\nu}_{\,\,\,\,p}\right)\right\}\nn\\
& &\qquad\left. -\frac{1}{12}(\b{\Delta})^{1/2}\left
\{e^{-\b {\phi}^{(1)}_D}
h_{\mu\nu\lambda}^{(1)} h^{(1)\,\mu\nu\lambda} + e^{\b {\phi}^{(1)}_D}
\left(h^{(2)}_{\mu\nu\lambda} +
\phi^{(2)} h^{(1)}_{\mu\nu\lambda}\right)\left(h^{(2)\,\mu\nu\lambda} +
\phi^{(2)} h^{(1)\,\mu\nu\lambda}\right)\right\}\right]\nn
\eqaend
where we have defined $\b {\phi}^{(1)}_D \,\,
\equiv\,\,\phi^{(1)}_D + \frac{1}
{2} \log \Delta$. Also, $g_{mn}\,\,=\,\,e^{\frac{4}{D-2} \phi^{(1)}_D}
\b {g}_{mn}$ and so, $\Delta\,\,=\,\,
e^{2\frac{(10-D)}{(D-2)} \phi^{(1)}_D}
\b {\Delta}$, with $\b \Delta^2\,\,
\equiv\,\,({\rm det}\,\b {g}_{mn})$. Note that
unlike $\h {\phi}^{(1)}$, $\b {\phi}^{(1)}_D$ is a D-dimensional field 
composed of D-dimensional scalars $\phi^{(1)}_D$ and $\Delta$. By defining
\begineq
{\cal M}_D\,\,\equiv\,\, \left(\begin{array}{cc}
\left( {\phi}^{(2)}\right)^2 e^{\b {\phi}^{(1)}_D} + 
e^{- \b {\phi}^{(1)}_D}
&  {\phi}^{(2)} e^{\b {\phi}^{(1)}_D} \\
 {\phi}^{(2)} e^{\b {\phi}^{(1)}_D}  &  e^{\b {\phi}^{(1)}_D}\end{array}\right)
\endeq
the action (18) can be expressed in a manifestly SL(2, R) invariant 
form\footnote[2]{This corrects the inference drawn in ref.[18] about
the SL(2, R) non-invariance of the D = 9 type IIB action.}
as follows:
\eqabegin
& &\int\,d^D x \sqrt{-\b {g}_D}\left[\b {R}_D + \frac{1}{4} {\rm tr}\,
\partial_\mu {\cal M}_D \partial^\mu {\cal M}_D^{-1} +
\frac{1}{8} \partial_\mu \log
\b {\Delta} \partial^\mu \log \b {\Delta} + \frac{1}{4} \partial_\mu
\b {g}_{mn} \partial^\mu \b {g}^{mn}\right.\nn\\
& &\qquad\qquad\qquad -\frac{1}{4} \b {g}_{mn} f_{\mu\nu}^{(3)\,m} f^{(3)\,
\mu\nu,\,n} - \frac{1}{4} (\b \Delta)^{1/2} \b {g}^{mp} \b {g}^{nq}
\partial_\mu {\bf b}_{mn}^T {\cal M}_D \partial^\mu {\bf b}_{pq}\\
& &\qquad\qquad\qquad\left. -\frac{1}{4} 
(\b \Delta)^{1/2} \b {g}^{mp} {\bf h}
^T_{\mu\nu\,m} {\cal M}_D {\bf h}^{\mu\nu}_{\,\,\,\,p} - \frac{1}{12}
(\b \Delta)^{1/2} {\bf h}_{\mu\nu\lambda}^T {\cal M}_D 
{\bf h}^{\mu\nu\lambda}\right]\nn
\eqaend
Here we have defined ${\bf b}_{mn}\,\,\equiv \,\,
 \left(\begin{array}{c} b_{mn}^{(1)}  \\
b_{mn}^{(2)}\end{array}\right)$, ${\bf h}_{\mu\nu\,m}\,\,\equiv \,\,
\left(\begin{array}{c} h_{\mu\nu\,m}^{(1)} \\ h_{\mu\nu\,m}^{(2)} \end{array}
\right)$, and ${\bf h}_{\mu\nu\lambda}\,\,\equiv\,\,\left(\begin{array}{c}
h_{\mu\nu\lambda}^{(1)} \\ h_{\mu\nu\lambda}^{(2)}\end{array}\right)$. 
Now it is clear that the 
action (20) is invariant under the following global SL(2, R) transformation: 
\eqabegin
{\cal M}_D &\rightarrow& \Lambda {\cal M}_D \Lambda^T, \qquad {\bf b}_{mn}
\,\,\,\rightarrow\,\,\,(\Lambda^{-1})^T {\bf b}_{mn}\nn\\ 
\left(\begin{array}{c} a_{\mu\,m}^{(1)}\\ a_{\mu\,m}^{(2)}\end{array}\right)
&\equiv& {\bf a}_{\mu\,m}\,\,\,\rightarrow\,\,\,(\Lambda^{-1})^T
{\bf a}_{\mu\,m},\qquad
\left(\begin{array}{c} b_{\mu\nu}^{(1)}\\ b_{\mu\nu}^{(2)}\end{array}\right)
\,\,\equiv\,\, {\bf b}_{\mu\nu}\,\,\,\rightarrow\,\,\,(\Lambda^{-1})^T 
{\bf b}_{\mu\nu}\nn\\
 \b {g}_{\mu\nu}&\rightarrow& \b {g}_{\mu\nu},
\qquad \b {g}_{mn}\,\,\,\rightarrow\,\,\,\b {g}_{mn},\qquad
{\rm and}\quad a_{\mu}^{(3)\,m}\,\,\,\rightarrow\,\,\,a_\mu^{(3)\,m}
\eqaend
where $\Lambda\,\,=\,\,\left(\begin{array}{cc} a & b\\ c & d\end{array}
\right)$, with $ad - bc = 1$, is the SL(2, R) transformation matrix. The
same transformation rules have been found in ref.[16] using different method.
Note, however, that it is the complex scalar $\rho_D = \phi^{(2)} + i
e^{-\b {\phi}_D^{(1)}}$ which undergoes the usual
fractional linear transformation under (21), where $\rho_D$ involves
$\b \phi_D^{(1)}$ and not $\phi_D^{(1)}$, the
D-dimensional dilaton. Also, note that it is $\b {\Delta}$ which remains 
invariant under SL(2, R) transformation, whereas $\Delta$ changes. This
follows from the fact that the internal metric $g_{mn}$ does not remain
invariant under
SL(2, R) transformation, whereas the scaled metric $\b {g}_{mn}$ is invariant.

Let us then look at the special case of this SL(2, R) transformation i.e.
the Z$_2$ subgroup generated by $\Lambda\,\,=\,\,\left(\begin{array}
{cc} 0 & 1\\ -1 & 0\end{array}\right)$. We have mentioned before for the
ten dimensional case that it generates the strong-weak coupling duality
when $\h {\phi}^{(2)} = 0$. For the D-dimensional action if we set $\phi^
{(2)} = 0$, the same $\Lambda$ generates the symmetry $\b {\phi}^{(1)}_D
\,\,\rightarrow\,\, - \b {\phi}^{(1)}_D$. Since $\b {\phi}^{(1)}_D$ is
 not the D-dimensional dilaton, this symmetry does not necessarily mean
a strong-weak coupling duality symmetry in the D-dimensional theory. By
demanding that
\begineq
\b {\Delta} = e^{-2 \frac{(10-D)}{(D-2)} \phi^{(1)}_D} \Delta
\endeq
remains invariant under Z$_2$ transformation, we find the transformation
rules for the D-dimensional dilaton $\phi^{(1)}_D$ and $\log \Delta$ as
follows:
\eqabegin
\phi^{(1)}_D &\rightarrow& \frac{6-D}{4}\phi^{(1)}_D + \frac{2-D}{8}
\log \Delta\nn\\
\log \Delta &\rightarrow& \frac{D-10}{2} \phi^{(1)}_D + \frac{D-6}{4}
\log \Delta
\eqaend
In particular, for D=10, we get from (23) $\phi^{(1)}_{10}
\rightarrow -\phi^{(1)}_{10}$ as expected. Also we note here that for 
D $\leq$ 6, $\phi^{(1)}_D$ does not change sign and so one might be tempted 
to think that in this case Z$_2$ transformation does not produce the
strong-weak coupling duality in the reduced theory. We point out that this
inference is not quite correct as $\log \Delta$ also changes under Z$_2$.
If we express $\b {\phi}^{(1)}_D$ in terms of SL(2, R) invariant quantity
$\b \Delta$, we find,
\begineq
\b {\phi}^{(1)}_D = \frac{8}{D-2} \phi^{(1)}_D + \frac{1}{2} \log 
\b {\Delta}
\endeq
Now demanding $\b {\phi}_D^{(1)} \rightarrow -\b {\phi}^{(1)}_D$ under
Z$_2$, we find the transformation rule for the D-dimensional dilaton
as
\begineq
\phi^{(1)}_D \rightarrow -\phi^{(1)}_D + \frac{2-D}{8} \log \b {\Delta}
\endeq
implying a strong-weak coupling duality in the D-dimensional theory apart
from a constant scaling factor $(\b {\Delta})^{(2-D)/8}$. We have thus 
clarified how the Z$_2$ subgroup of SL(2, R) group in the ten-dimensional
theory induces a strong-weak coupling duality in the reduced theory.

Next we show that the SL(2, R) symmetry of the type IIB string theory
is not only the symmetry of the truncated action (1), but also is a
symmetry of the full theory compactified on torus. As the full type IIB
action is not known we take the complete type IIA string effective action [18]
and then compactify on a circle. We then make a T-duality 
transformation [24,18]
on this 9D action to convert it into type IIB action. This T-dual action
will be shown to have the SL(2, R) invariance. The massless spectrum of the
bosonic sector of type IIA string theory consists of a metric $\h {g}_{\h \mu
\h \nu}$, an antisymmetric tensor field $\h {B}_{\h \mu \h \nu}^{(1)}$ and
a dilaton $\h \phi$ in the NS-NS sector and in the R-R sector it has a vector
gauge field $\h {A}^{(1)}_{\h \mu}$ and a 
three-form antisymmetric tensor field
$\h {C}_{\h \mu \h \nu \h \lambda}$
[21]. The full action in D=10 is given in the
following [18]:
\eqabegin
S_{{\rm IIA}}^{(10)} &=& \int\,d^{10} \h x \sqrt{- \h g}\left[e^{-2 \h \phi}
\left(\h R + 4 \partial_{\h \mu} \h \phi \partial^{\h \mu} \h \phi - \frac
{1}{12} \h H_{\h \mu \h \nu \h \lambda}^{(1)} \h H^{(1)\,\h \mu \h \nu \h
\lambda}\right) - \frac{1}{4} \h F_{\h \mu \h \nu}^{(1)} 
\h F^{(1)\,\h \mu \h \nu}\right.
\nn\\
& &\qquad\qquad -\frac{1}{12}\h F_{\h \mu \h \nu \h \lambda \h \rho}
\h F^{\h \mu \h \nu \h \lambda \h \rho} + \frac{4}{(12)^4} \frac{\epsilon
^{\h \mu_1 \ldots \h \mu_{10}}}{\sqrt{ - \h g}}\left(3 \h F_{\h \mu_1 \ldots
\h \mu_4} \h F_{\h \mu_5 \ldots \h \mu_8} \h B_{\h \mu_9 \h \mu_{10}}^{(1)}
\right.\nn\\
& &\qquad\qquad\qquad\qquad\left.\left. - 8 \h F_{\h \mu_1 \ldots \h \mu_4} 
\h H_{\h \mu_5
\h \mu_6 \h \mu_7} \h C_{\h \mu_8 \h \mu_9 \h \mu_{10}}\right)\right]
\eqaend
where $\h H_{\h \mu \h \nu \h \lambda}^{(1)} = \left(\partial_{\h \mu}
\h B_{\h \nu \h \lambda}^{(1)} + {\rm cyc.\,\, in\,\,} \h \mu \h \nu \h 
\lambda\right),\,\, \h F_{\h \mu \h \nu}^{(1)} = \partial_{\h \mu} \h
A_{\h \nu}^{(1)} - \partial_{\h \nu} \h A_{\h \mu}^{(1)}$ and $\h F_{\h
\mu \h \nu \h \lambda \h \rho} = \partial_{\h \mu} \h C_{\h \nu \h \lambda
\h \rho} - \partial_{\h \nu} \h C_{\h \mu \h \lambda \h \rho} + \partial_
{\h \lambda} \h C_{\h \rho \h \mu \h \nu} - \partial_{\h \rho} \h C_{\h
\lambda \h \mu \h \nu} + \left(\h F_{\h \mu \h \nu}^{(1)} \h B_{\h \lambda 
\h \rho}^{(1)} + \h F_{\h \nu \h \lambda}^{(1)} \h B_{\h \mu \h \rho}^{(1)}
+ {\rm cyc.\,\, in\,\,} \h \nu \h \lambda \h \rho\right)$.
Note that the first three terms in (26) i.e. the NS-NS terms couple to the
dilaton, whereas, the fourth and the fifth terms are R-R terms and do not
couple to the dilaton. But all these terms are dependent on the metric 
$\h g_{\h \mu \h \nu}$. On the other hand, the last two terms are mixed 
terms and do not depend on the metric. In this sense these terms are 
topological. The reduced forms of these terms depend on the dimensionality
of the theory. The reduction of this action (26) on a circle has already
been performed in ref.[18]. We here write the D=9 type IIA string effective 
action without the topological terms (the SL(2, R) invariance of the T-dual
form of the topological terms will be considered later),
\eqabegin
S_{{\rm IIA}}^{(9)\,'} &=& \int\,d^9 x \sqrt{-g}\left[e^{-2\phi}\left(
R + 4 \partial_\mu \phi \partial^\mu\phi - \frac{1}{4} \partial_\mu 
\log \chi \partial^\mu \log \chi -\frac{1}{4} \chi F_{\mu\nu}^{(2)}
F^{(2)\,\mu\nu}\right.\right.\nn\\
& &\left. - \frac{1}{4} \chi^{-1} F_{\mu\nu}^{(3)} F^{(3)\,\mu\nu}
-\frac{1}{12}H_{\mu\nu\lambda}^{(1)} H^{(1)\,\mu\nu\lambda}\right) - \frac
{1}{2} \chi^{-\frac{1}{2}} \partial_\mu \psi \partial^\mu \psi 
- \frac{1}{12} \chi
^{\frac{1}{2}} F_{\mu\nu\lambda\rho} F^{\mu\nu\lambda\rho}\nn\\
& & -\frac{1}{12} \chi^{-\frac{1}{2}}\left(H_{\mu\nu\lambda}^{(2)} -
\psi H_{\mu\nu\lambda}^{(1)}\right)\left(H^{(2)\,\mu\nu\lambda} - \psi
H^{(1)\,\mu\nu\lambda}\right)\nn\\ 
& &\qquad\qquad\left. - \frac{1}{4} \chi^{\frac{1}{2}}
\left(F_{\mu\nu}
^{(1)} + \psi F_{\mu\nu}^{(2)}\right)\left(F^{(1)\,\mu\nu} + \psi F^{(2)\,
\mu\nu}\right)\right]
\eqaend
where the definitions of various reduced fields and their field-strengths
are listed below:
\eqabegin
\h g_{\h \mu \h \nu} &\longrightarrow& \left\{ \begin{array}{l} \h g_{99}
= g_{99} = \chi
\\
\h g_{\mu 9} = g_{\mu 9} = \chi A_\mu ^{(2)}\\
\h g_{\mu \nu} = g_{\mu\nu} + \chi A_\mu ^{(2)} A_\nu^{(2)} 
\end{array}\right.\\
\h \phi &=& \phi + \frac{1}{4} \log \chi\\
\h A_{\h \mu}^{(1)} &\longrightarrow& \left\{\begin{array}{l}
A_9 = \h A_9 = \psi \\
A_\mu^{(1)} = \h A_{\mu}^{(1)} - \psi A_\mu^{(2)}
\end{array}\right.\\
\h B_{\h \mu \h \nu}^{(1)} &\longrightarrow& \left\{\begin{array}{l}
B_{\mu 9}^{(1)} = \h B_{\mu 9}^{(1)} = A_\mu^{(3)} \\
B_{\mu\nu}^{(1)} = \h B_{\mu\nu}^{(1)} + A_\mu^{(2)} A_\nu^{(3)} -
A_\nu^{(2)} A_\mu^{(3)}\end{array}\right.\\
\h C_{\h \mu \h \nu \h \lambda} &\longrightarrow& \left\{\begin{array}{l}
C_{\mu\nu 9} = \h C_{\mu\nu 9} = B_{\mu\nu}^{(2)} - \psi B_{\mu\nu}^{(1)}
- \left(A_\mu^{(1)} A_\nu^{(3)} - A_\nu^{(1)} A_\mu^{(3)}\right)\\
C_{\mu\nu\lambda} = \h C_{\mu\nu\lambda} - \left(A_\mu^{(2)} \h C_
{\nu\lambda 9} + {\rm cyc.\,\,in\,\,} \mu\nu\lambda\right)
\end{array}\right.
\eqaend
and the field strengths are:
\eqabegin
F_{\mu\nu}^{(2)} &=& \partial_\mu A_\nu^{(2)} - \partial_\nu A_\mu^{(2)},
\qquad F_{\mu\nu}^{(3)}\,\,\,=\,\,\,\partial_\mu A_\nu^{(3)} - \partial_
\nu A_\mu^{(3)},\qquad F_{\mu 9}\,\,\,=\,\,\, \partial_\mu \psi\nn\\
F_{\mu\nu} &=& F_{\mu\nu}^{(1)} + \psi F_{\mu\nu}^{(2)},\qquad
H_{\mu\nu 9}^{(1)}\,\,\,=\,\,\,F_{\mu\nu}^{(3)}\\
H_{\mu\nu\lambda}^{(1)} &=& \partial_\mu B_{\nu\lambda}^{(1)} - 
F_{\mu\nu}^{(2)} A_\lambda^{(3)} + {\rm cyc.\,\,in\,\,\mu\nu\lambda}\nn\\
&=&\partial_\mu \b {B}_{\nu\lambda}^{(1)} - \frac{1}{2}\left(F_{\mu\nu}
^{(2)} A_\lambda^{(3)} + F_{\mu\nu}^{(3)} A_\lambda^{(2)}\right) +
{\rm cyc.\,\,in\,\,\mu\nu\lambda}
\eqaend
we have defined
\begineq
\b {B}_{\mu\nu}^{(1)} \equiv B_{\mu\nu}^{(1)} - \frac{1}{2}\left(A_\mu^{(2)}
A_\nu^{(3)} - A_\nu^{(2)} A_\mu^{(3)}\right)
\endeq
Continuing with other field strengths,
\begineq
F_{\mu\nu\lambda 9} = \h F_{\mu\nu\lambda 9} = H_{\mu\nu\lambda}^{(2)}
- \psi H_{\mu\nu\lambda}^{(1)}
\endeq
where $H_{\mu\nu\lambda}^{(2)}$ is defined as,
\eqabegin
H_{\mu\nu\lambda}^{(2)} &=& \partial_\mu B_{\nu\lambda}^{(2)} + 
F_{\mu\nu}^{(3)} A_\lambda^{(1)} + {\rm cyc.\,\,in\,\,} \mu\nu\lambda\nn\\
&=& \partial_\mu \b {B}_{\nu\lambda}^{(2)} + \frac{1}{2}\left(F_{\mu\nu}
^{(1)} A_\lambda^{(3)} + F_{\mu\nu}^{(3)} A_\lambda^{(1)}\right) + {\rm
cyc.\,\,in\,\,} \mu\nu\lambda
\eqaend
where
\begineq
\b {B}_{\mu\nu}^{(2)} \equiv B_{\mu\nu}^{(2)} - \frac{1}{2}\left(A_\mu^{(1)}
A_\nu^{(3)} - A_\nu^{(1)} A_\mu^{(3)}\right)
\endeq
and finally,
\eqabegin
F_{\mu\nu\lambda\rho} &=& \partial_\mu \b C_{\nu\lambda\rho} - \partial_\nu
\b C_{\mu\lambda\rho} + \partial_\lambda \b C_{\rho\mu\nu} - \partial_\rho
\b C_{\lambda\mu\nu} + \left[F_{\mu\nu}^{(i)} \b B_{\lambda\rho}^{(i)}
+ F_{\nu\lambda}^{(i)} \b B_{\mu\rho}^{(i)}\right.\nn\\
& & \left. -\frac{1}{2} \epsilon^{ij} F_{\mu\nu}^{(3)} A_\lambda^{(i)}A_\rho
^{(j)} - \frac{1}{2} \epsilon^{ij}F_{\nu\lambda}^{(3)}A_\mu^{(i)}A_\rho^{(j)}
+ {\rm cyc.\,\,in\,\,} \nu\lambda\rho\right]
\eqaend
where
\begineq
\b C_{\mu\nu\lambda} = C_{\mu\nu\lambda} + \left[\frac{1}{2}\epsilon^{ij}
A_\mu^{(i)} A_\nu^{(j)} A_\lambda^{(3)} + {\rm cyc.\,\,in\,\,} \mu\nu
\lambda\right]
\endeq
Here $i,\,j = 1,\,2$ and $\epsilon^{12}\,=\,-\epsilon^{21}\,=\,1$.
Note here that we have introduced $\b B_{\mu\nu}^{(i)}$
fields. When we take T-duality transformation on the action (27), we will
use the fact that $\b B_{\mu\nu}^{(1)}$ does not transform whereas $\b B_{
\mu\nu}^{(2)}$ transforms in a non-local way. The T-duality transformation
for which the NS-NS sector of the action (27) remains invariant is given by
[18]:
\begineq
\tilde \chi = \chi^{-1}, \quad \tilde {A}_\mu^{(2)} = - A_\mu^{(3)},
\quad \tilde {A}_\mu^{(3)} = - A_\mu^{(2)}, \quad \tilde{g}_{\mu\nu}
= g_{\mu\nu},\quad \tilde \phi = \phi
\endeq
We will assume that both $H_{\mu\nu\lambda}^{(1)}$ and 
$H_{\mu\nu\lambda}^{(2)}$ do not transform under T-duality transformation.
For $H^{(1)}$, it is clear from (34) and (41) that it is indeed invariant if 
$\b B_{\mu\nu}^{(1)}$ is invariant, whereas, for $H^{(2)}$, we note from
(37) that it will remain invariant if $\b B_{\mu\nu}^{(2)}$ transforms in
an appropriate non-local way. We will also assume that 
$F_{\mu\nu\lambda\rho}$ remains
invariant under T-duality transformation, but it means from (39) that
$\b C_{\mu\nu\lambda}$ should transform appropriately. We should emphasize 
that these assumptions are made just for simplification. Now with
these T-duality rules the action (27) changes as:
\eqabegin
\tilde {S}_{{\rm IIA}}^{(9)\,'} &=& S_{{\rm IIB}}^{(9)\,'}\nn\\
&=& 
\int\,d^9 x \sqrt{-g}\left[e^{-2\phi}\left(
R + 4 \partial_\mu \phi \partial^\mu\phi - \frac{1}{4} \partial_\mu 
\log \chi \partial^\mu \log \chi -\frac{1}{4} \chi F_{\mu\nu}^{(2)}
F^{(2)\,\mu\nu}\right.\right.\nn\\
& &\left. - \frac{1}{4} \chi^{-1} F_{\mu\nu}^{(3)} F^{(3)\,\mu\nu}
-\frac{1}{12}H_{\mu\nu\lambda}^{(1)} H^{(1)\,\mu\nu\lambda}\right) - \frac
{1}{2} \chi^{\frac{1}{2}} \partial_\mu \psi \partial^\mu \psi 
- \frac{1}{12} \chi
^{-\frac{1}{2}} F_{\mu\nu\lambda\rho} F^{\mu\nu\lambda\rho}\nn\\
& & -\frac{1}{12} \chi^{\frac{1}{2}}\left(H_{\mu\nu\lambda}^{(2)} -
\psi H_{\mu\nu\lambda}^{(1)}\right)\left(H^{(2)\,\mu\nu\lambda} - \psi
H^{(1)\,\mu\nu\lambda}\right)\nn\\
& &\qquad\qquad\left.- \frac{1}{4} \chi^{-\frac{1}{2}}
\left(F_{\mu\nu}
^{(1)} - \psi F_{\mu\nu}^{(3)}\right)\left(F^{(1)\,\mu\nu} - \psi F^{(3)\,
\mu\nu}\right)\right]
\eqaend
By rescaling the metric $g_{\mu\nu} = e^{\frac{4}{7} \phi} \b g_{\mu\nu}$
and $\chi = e^{\frac{4}{7} \phi} \b \chi$, we convert this part of the action
(42) in the Einstein frame as follows:
\eqabegin
& &\int\,d^9 x \sqrt{-\b g}\left(
\b R -\frac{1}{2} \partial_\mu \b \phi \partial^\mu \b \phi - \frac{1}{2}
e^{2 \b \phi} \partial_\mu \psi \partial^\mu \psi
- \frac{7}{32} \partial_\mu 
\log \b \chi \partial^\mu \log \b \chi\right.\nn\\
& &\qquad\qquad -\frac{1}{4}\b \chi F_{\mu\nu}^{(2)}
F^{(2)\,\mu\nu} -\frac{1}{12} \b \chi^{-1/2} F_{\mu\nu\lambda\rho} F^{\mu\nu
\lambda\rho}\nn\\
& &\qquad - \frac{1}{4}\b \chi^{-3/4} \left\{e^{-\b \phi}
F_{\mu\nu}^{(3)} F^{(3)\,\mu\nu} + e^{\b \phi}\left(F_{\mu\nu}^{(1)} -\psi
F_{\mu\nu}^{(3)}\right)\left(F^{(1)\,\mu\nu} - \psi F^{(3)\,\mu\nu}\right)
\right\}\nn\\
& &\quad\left. -\frac{1}{12}\b \chi^{1/4} \left\{e^{-\b \phi}
H_{\mu\nu\lambda}^{(1)} H^{(1)\,\mu\nu\lambda} +
e^{\b \phi}\left(H_{\mu\nu\lambda}^{(2)} -
\psi H_{\mu\nu\lambda}^{(1)}\right)\left(H^{(2)\,\mu\nu\lambda} - \psi
H^{(1)\,\mu\nu\lambda}\right)\right\}\right) 
\eqaend
Now by defining
\begineq
 {\cal N}_D\,\,\equiv\,\, \left(\begin{array}{cc}
\psi^2 e^{\b {\phi}} + 
e^{- \b {\phi}}
&  -\psi e^{\b {\phi}} \\
 -\psi e^{\b {\phi}}  &  e^{\b {\phi}}\end{array}\right)
\endeq
the part of the action (43) can be written in a manifestly SL(2, R) invariant
form as given below:
\eqabegin
& & \int\,d^9 x \sqrt{-\b g}\left(\b R + \frac{1}{4} {\rm tr}\,\partial_\mu
{\cal N}_D \partial^\mu {\cal N}_D^{-1} - \frac{7}{32} \partial_\mu \log{\b
\chi} \partial^\mu \log {\b \chi} - \frac{1}{4} \b \chi F_{\mu\nu}^{(2)}
F^{(2)\,\mu\nu}\right.\nn\\
& &\qquad \left.-\frac{1}{4} \b \chi^{-3/4} {\cal F}_{\mu\nu}^T {\cal N}_D {\cal F}
^{\mu\nu} -\frac{1}{12} \b \chi^{1/4} {\cal H}_{\mu\nu\lambda}^T {\cal N}_D
{\cal H}^{\mu\nu\lambda} - \frac{1}{12} \b \chi^{-1/2}F_{\mu\nu\lambda\rho}
F^{\mu\nu\lambda\rho}\right)
\eqaend
where we have defined ${\cal F}_{\mu\nu}\,\,\equiv\,\,\left(\begin{array}{l}
F_{\mu\nu}^{(3)} \\ F_{\mu\nu}^{(1)}\end{array}\right)$ and ${\cal H}_{\mu
\nu\lambda}\,\,\equiv\,\,\left(\begin{array}{l} H_{\mu\nu\lambda}^{(1)} \\
H_{\mu\nu\lambda}^{(2)}\end{array}\right)$.
So, the part of the action without the topological term is indeed invariant
under the following global SL(2, R) transformation:
\eqabegin
{\cal N}_D &\rightarrow& \Lambda {\cal N}_D \Lambda^T, \qquad \b g_{\mu\nu}
\,\,\rightarrow\,\,\b g_{\mu\nu},\qquad \b \chi\,\,\rightarrow\,\,\b \chi,
\qquad A_\mu^{(2)}\,\,\rightarrow\,\,A_\mu^{(2)}\nn\\
\left(\begin{array}{c} A_{\mu}^{(3)}\\ A_{\mu}^{(1)}\end{array}\right)
&\equiv& {\cal A}_{\mu}\,\,\,\rightarrow\,\,\,(\Lambda^{-1})^T
{\cal A}_{\mu}, \qquad
{\cal H}_{\mu\nu\lambda}\,\,\,\rightarrow\,\,\,
(\Lambda^{-1})^T 
{\cal H}_{\mu\nu\lambda}\nn\\
F_{\mu\nu\lambda\rho}
&\rightarrow& F_{\mu\nu\lambda\rho}
\eqaend
Note also that here $A_\mu^{(1)}$, 
$A_\mu^{(2)}$, $A_\mu^{(3)}$ play the same role as $a_\mu^{(2)}$,  
$a_\mu^{(3)}$ and $a_\mu^{(1)}$ respectively
in the truncated version of type IIB theory  as discussed earlier and 
so, those fields
can be identified. Similarly, $H_{\mu\nu\lambda}^{(1)}$, $H_{\mu\nu\lambda}
^{(2)}$ fields play the same role as $h_{\mu\nu\lambda}^{(1)}$, 
$h_{\mu\nu\lambda}^{(2)}$ in type IIB theory.

We now look at the topological part of the action (26) and show that this 
part is also SL(2, R) invariant. The dimensional reduction of the topological
part on a circle has also been obtained in ref.[18] and the explicit form
is given below:
\eqabegin
S_{{\rm IIA}}^{(9)\,''} &=& \int\,d^9 x \frac{1}{2 (12)^3} \epsilon^{\mu_1
\ldots \mu_9}\left(F_{\mu_1\ldots \mu_4} F_{\mu_5 \ldots \mu_8} A_{\mu_9}
^{(3)}
 -4 F_{\mu_1 \ldots \mu_4} \epsilon^{ij} H_{\mu_5\mu_6\mu_7}^{(i)}
\b {B}_{\mu_8\mu_9}^{(j)}\right.\nn\\ 
& &+ 4 F_{\mu_1 \ldots \mu_4} H_{\mu_5\mu_6\mu_7}
^{(i)} A_{\mu_8}^{(i)} A_{\mu_9}^{(3)} + 2 \epsilon^{ij} H_{\mu_1\mu_2\mu_3}
^{(i)} H_{\mu_4\mu_5\mu_6}^{(j)} \b {C}_{\mu_7\mu_8\mu_9}
 + 4 F_{\mu_1 \ldots \mu_4} F_{\mu_5\mu_6}^{(3)} \b {C}_
{\mu_7\mu_8\mu_9}\nn\\ 
& &\qquad\left.+ 6 \epsilon^{ij} H_{\mu_1\mu_2\mu_3}^{(i)} H_{\mu_4
\mu_5\mu_6}^{(j)} A_{\mu_7}^{(k)} \b {B}_{\mu_8\mu_9}^{(k)} + 12 
F_{\mu_1 \ldots \mu_4} F_{\mu_5\mu_6}^{(3)} A_{\mu_7}^{(i)} 
\b {B}_{\mu_8
\mu_9}^{(i)}\right)
\eqaend
We then make, as before, the T-duality transformation (41) and the rules
given afterwards to convert the topological part of the type IIA action
to type IIB action as follows:
\eqabegin
& &\tilde {S}_{{\rm IIA}}^{(9)\,''} \,\,\,=\,\,\, S_{{\rm IIB}}
^{(9)\,''}\nn\\
&=& \int\,d^9 x \frac{1}{2 (12)^3} \epsilon^{\mu_1 \ldots \mu_9}\left[
- F_{\mu_1 \ldots \mu_4} F_{\mu_5 \ldots \mu_8} A_{\mu_9}^{(2)}
- F_{\mu_1 \ldots \mu_4} H_{\mu_5\mu_6\mu_7}^{(1)} \tilde {\b B}_{\mu_8
\mu_9}^{(2)}\right.\nn\\
&+& F_{\mu_1 \ldots \mu_4} H_{\mu_5\mu_6\mu_7}^{(2)} \b 
{B}_{\mu_8\mu_9}^{(1)} - 4 F_{\mu_1 \ldots \mu_4} H_{\mu_5\mu_6\mu_7}^{(1)}
A_{\mu_8}^{(1)} A_{\mu_9}^{(2)} + 4 F_{\mu_1 \ldots \mu_4} H_
{\mu_5\mu_6\mu_7}^{(2)} A_{\mu_8}^{(3)} A_{\mu_9}^{(2)}\nn\\
&+&  2 H_{\mu_1\mu_2\mu_3}^{(1)} H_{\mu_4\mu_5\mu_6}^{(2)} \tilde {\b C}
_{\mu_7\mu_8\mu_9} - 2 H_{\mu_1\mu_2\mu_3}^{(2)} H_{\mu_4\mu_5\mu_6}^{(1)}
\tilde {\b C}_{\mu_7\mu_8\mu_9} - 4 F_{\mu_1 \ldots \mu_4} F_{\mu_5\mu_6}
^{(2)} \tilde {\b C}_{\mu_7\mu_8\mu_9}\nn\\
&+& 6\left(H_{\mu_1\mu_2\mu_3}^{(1)} H_{\mu_4\mu_5\mu_6}^{(2)} 
A_{\mu_7}^{(1)}
\b {B}_{\mu_8\mu_9}^{(1)} - H_{\mu_1\mu_2\mu_3}^{(1)} H_{\mu_4\mu_5\mu_6}
^{(2)} A_{\mu_7}^{(3)} \tilde {\b B}_{\mu_8\mu_9}^{(2)} - H_{\mu_1
\mu_2\mu_3}^{(2)} H_{\mu_4\mu_5\mu_6}^{(1)} A_{\mu_7}^{(1)} \b {B}_{\mu_8
\mu_9}^{(1)}\right.\nn\\
&+& \left.\left.  H_{\mu_1\mu_2\mu_3}^{(2)} H_{\mu_4\mu_5\mu_6}^{(1)} 
A_{\mu_7}^{(3)}
\tilde {\b B}_{\mu_8\mu_9}^{(2)}\right) - 12\left(F_{\mu_1 
\ldots \mu_4} F_{\mu_5
\mu_6}^{(2)} A_{\mu_7}^{(1)} \b {B}_{\mu_8\mu_9}^{(1)} + F_{\mu_1
\ldots \mu_4} F_{\mu_5\mu_6}^{(2)} A_{\mu_7}^{(3)} \tilde {\b B}_{\mu_8\mu_9}
^{(2)}\right)\right]\nn\\
\eqaend
Here, $\tilde {\b B}_{\mu\nu}^{(2)}$ and $\tilde {\b C}_{\mu\nu\lambda}$
denote the T-dual forms of the corresponding fields $\b B_{\mu\nu}^{(2)}$,
$\b C_{\mu\nu\lambda}$ whose explicit forms are not important. If we now
introduce SL(2, R) metric $\Sigma = \left(\begin{array}{cc} 0 & i \\
-i & 0 \end{array}\right)$ which satisfies $\Lambda \Sigma \Lambda^T
= \Sigma$ and $\Sigma \Lambda \Sigma = \Lambda^{-1}$ for any SL(2, R)
matrix $\Lambda$, then in terms of $\Sigma$, (48) can be written in a 
manifestly SL(2, R) invariant form as given below:
\eqabegin
& & \int\,d^9 x \frac{1}{2 (12)^3} \epsilon^{\mu_1 \ldots \mu_9}\left(
- F_{\mu_1 \ldots \mu_4} F_{\mu_5 \ldots \mu_8} A_{\mu_9}^{(2)} - i
F_{\mu_1 \ldots \mu_4} {\cal H}_{\mu_5\mu_6\mu_7}^T \Sigma {\cal B}
_{\mu_8\mu_9}\right.\nn\\
& & - 4i F_{\mu_1 \ldots \mu_4} {\cal H}_{\mu_5\mu_6\mu_7}^T \Sigma 
{\cal A}_{\mu_8} A_{\mu_9}^{(2)} + 2i {\cal H}_{\mu_1\mu_2\mu_3}^T
\Sigma {\cal H}_{\mu_4\mu_5\mu_6} \tilde {\b C}_{\mu_7\mu_8\mu_9} 
- 4 F_{\mu_1 \ldots \mu_4} F_{\mu_5\mu_6}^{(2)} \tilde {\b C}_{\mu_7
\mu_8\mu_9}\nn\\
& &\left. + 6 {\cal H}_{\mu_1\mu_2\mu_3}^T \Sigma {\cal H}_{\mu_4\mu_5
\mu_6} {\cal A}_{\mu_7}^T \Sigma {\cal B}_{\mu_8\mu_9} + 12i 
F_{\mu_1 \ldots \mu_4} F_{\mu_5\mu_6}^{(2)} {\cal A}_{\mu_7}^T \Sigma
{\cal B}_{\mu_8\mu_9}\right)
\eqaend
Thus it is clear that the T-dual form of the topological part of the 
action (26) compactified on a circle is SL(2, R) invariant under the 
transformations (46) alongwith
\begineq
\tilde {\b C}_{\mu\nu\lambda} \rightarrow \tilde {\b C}_{\mu\nu\lambda}
\endeq
Note that ${\cal H}_{\mu\nu\lambda}$ transformation in (46) implies
that ${\cal B}_{\mu\nu}\,\,\equiv\,\,\left(\begin{array}{l} \b B_{\mu\nu}
^{(1)} \\ \tilde {\b B}_{\mu\nu}^{(2)}\end{array}\right)$ transforms
as ${\cal B}_{\mu\nu} \rightarrow (\Lambda^{-1})^T {\cal B}_{\mu\nu}$.
So, we have shown that the T-dual form of the full type IIA action in 9D
i.e. the full type IIB string effective action in 9D is invariant under the
SL(2, R) transformations (46) and (50).

To conclude, we have studied the toroidal compactification of a truncated
version (when the self-dual five-form field strength is zero) of the type
IIB string effective action in the string frame. As we finally converted
the reduced action in the Einstein frame by conformal rescaling of the 
metric we have recovered the SL(2, R) invariance of the reduced action as
a consequence of the same symmetry in ten dimensions. We have obtained the
transformation properties of the various fields and compared with the 
recently obtained results of toroidal compactification of the same type IIB
action in the Einstein frame. Since the SL(2, R) matrix ${\cal M}_D$ in
eq.(19) involved in the process of showing the invariance does not contain
the D-dimensional dilaton, the issue of strong-weak coupling duality symmetry
under a Z$_2$ subgroup of this SL(2, R) group becomes confusing. We have 
clarified this point and have shown how the Z$_2$ subgroup produces the
strong-weak coupling duality in the reduced theory. We have also shown how
SL(2, R) invariance can be understood in the full type IIB string theory.
In the absence of the complete type IIB action, we have taken a T-dual
version of the complete type IIA action compactified on a circle. In this
simplest case, we were able to show that the full action is indeed SL(2, R)
invariant. It is quite involved to study the general case because of the
complication of the topological terms whose forms are dependent on the
dimensionality of the reduced theory. But, the invariance in 9D suggests that
the full type IIB string effective action is SL(2, R) invariant.  
Using this symmetry it will be interesting to find the 
SL(2, Z) multiplets of various classical 
p-brane solutions of lower dimensional type II string theory.
\vskip 1cm
\begin{large}
\noindent{\bf References:}
\end{large}

\vskip .5cm

\begin{enumerate}
\item C. Hull and P. Townsend, \np 438 (1995) 109.
\item C. Hull, \pl 357 (1995) 545.
\item J. H. Schwarz, \pl 360 (1995) 13; E: \pl 364 (1995) 252; {\it Superstring
Dualities}, hep-th/9509148.
\item A. Font, L. Ibanez, D. Lust and F. Quevedo, \pl 249 (1990) 35; S. J. Rey,
Phys. Rev D43 (1991) 35.
\item A. Sen, Int. J. Mod. Phys. A9 (1994) 3707; J. H. Schwarz, {\it String
Theory Symmetries}, hep-th/9503127.
\item J. H. Schwarz, \pl 367 (1996) 97.
\item J. H. Schwarz, {\it Lectures on Superstring and M-Theory Dualities},
hep-th/9607201.
\item E. Witten, \np 460 (1996) 335.
\item A. Dabholkar, {\it Ten Dimensional Heterotic String as a Soliton},
hep-th/9506160; A. Sen, {\it Unification of String Dualities}, hep-th/9609176.
\item P. Aspinwall, {\it Some Relationships Between Dualities in String
Theory}, in Proc. Trieste Conf. on ``S-Duality and Mirror Symmetry'', 
hep-th/9508154.
\item C. Vafa, \np 469 (1996) 403.
\item J. Scherk and J. Schwarz, \np 153 (1979) 61.
\item E. Cremmer, in {\it Supergravity}, 1981 eds. S. Ferrara and J. G. Taylor
(Cambridge University Press, 1982).
\item J. H. Schwarz, \np 226 (1983) 269.
\item P. Howe and P. West, \np 238 (1984) 181.
\item J. Maharana, {\it S-Duality and Compactification of Type IIB Superstring
Action}, hep-th/9703009.
\item E. Bergshoeff, C. M. Hull and T. Ortin, \np 451 (1995) 547.
\item A. Das and S. Roy, \np 482 (1996) 119.
\item M Dine , P. Huet and N. Seiberg, \np 322 (1989) 301.
\item J. Dai, R. G. Leigh and J. Polchinski, Mod. Phys. Lett. A4 (1989) 2073.
\item M. B. Green, J. Schwarz and E. Witten, {\it Superstring Theory}, Vols.
I and II, Cambridge University Press, 1987.
\item E. Cremmer and B. Julia, \np 159 (1979) 141.
\item J. Maharana and J. Schwarz, \np 390 (1993) 3; S. F. Hassan and A. Sen,
\np 375 (1992) 103.
\item T. Buscher, \pl 194 (1987) 59; \pl 201 (1988) 466.

\end{enumerate}

\vfil
\eject 

\end{document}